\documentclass[aps,prb,superscriptaddress,twocolumn, 10pt]{revtex4-1}
\usepackage{graphicx}
\usepackage{bm}
\usepackage[usenames]{color}
\bibstyle{apsrev.bib}

\usepackage{epsfig}
\usepackage{amsmath}
\DeclareMathOperator{\Tr}{Tr}
\usepackage{amssymb}
\usepackage{array}
 
\newcommand{\be}{\begin{equation}}
\newcommand{\ee}{\end{equation}}
\newcommand{\beqn}{\begin{eqnarray}}
\newcommand{\eeqn}{\end{eqnarray}}

\usepackage{dsfont}

\begin{document}

\title{Corrections to the singlet-length distribution and the entanglement entropy in random singlet phases}

\author{R\'obert Juh\'asz}
\email{juhasz.robert@wigner.hu}
\affiliation{Wigner Research Centre for Physics, Institute for Solid State Physics and Optics, H-1525 Budapest, P.O. Box 49, Hungary}

\date{\today}

\begin{abstract}
We consider random singlet phases of spin-$\frac{1}{2}$, random, antiferromagnetic spin chains, in which the universal leading-order divergence $\frac{\ln 2}{3}\ln\ell$ of the average entanglement entropy of a block of $\ell$ spins, as well as the closely related leading term $\frac{2}{3}l^{-2}$ in the distribution of singlet lengths are well known by the strong-disorder renormalization group (SDRG) method. Here, we address the question of how large the subleading terms of the above quantities are. 
By an analytical calculation performed along a special SDRG trajectory of the random XX chain, we identify a series of integer powers of $1/l$ in the singlet-length distribution with the subleading term $\frac{4}{3}l^{-3}$. Our numerical SDRG analysis shows that, for the XX fixed point, the subleading term is generally $O(l^{-3})$ with a non-universal coefficient and also reveals terms with half-integer powers: $l^{-7/2}$ and $l^{-5/2}$ for the XX and XXX fixed points, respectively. We also present how the singlet lengths originating in the SDRG approach can be interpreted and calculated in the XX chain from the one-particle states of the equivalent free-fermion model. 
These results imply that the subleading term next to the logarithmic one in the entanglement entropy is $O(\ell^{-1})$ for the XX fixed point and $O(\ell^{-1/2})$ for the XXX fixed point with non-universal coefficients. For the XX model, where a comparison with exact diagonalization is possible, the order of the subleading term is confirmed but we find that the SDRG fails to provide the correct non-universal coefficient.  
\end{abstract}

\maketitle

%%%%%%%%%%%%%%%%%%%%%%%%%%%%%%%%%%%%%%%%%%%%%%%%%%%%%%%%%%%%%%%%%%%%%%%%%%%%%
%%%%%%%%%%%%%%%%%%%%%%%%%%%%%%%%%%%%%%%%%%%%%%%%%%%%%%%%%%%%%%%%%%%%%%%%%%%%%

\section{Introduction}

The entanglement properties of critical quantum many-body systems have attracted much interest recently \cite{amico,eisert2010,entanglement_review,laflorencie_rev,holzhey,vidal,Calabrese_Cardy04, hsu2009}.  
A frequently used measure of quantum entanglement is the so-called entanglement entropy \cite{horodecki}. Considering a system in a pure state $| \Psi \rangle$, and a subsystem $A$, the entanglement entropy of $A$ is defined as the von Neumann entropy of the reduced density matrix $\rho_A$:
\be
S_A = -\Tr_A \rho_A \ln \rho_A.
\ee
Here, $\rho_A$ is obtained as a partial trace over the degrees of freedom of the environment $\overline{A}$: $\rho_{A} = \Tr_{\overline{A}} | \Psi \rangle \langle \Psi |$.  

In one-dimensional, infinite, translationally invariant, critical systems, conform field theory predicts a logarithmic asymptotic dependence of the entanglement entropy on the size $\ell$ of the subsystem as 
\be 
S_{\ell}=\frac{c}{3}\ln\ell + {\rm const},
\ee
where $c$ is the central charge of the conformal algebra \cite{holzhey,vidal,Calabrese_Cardy04}.

Besides homogeneous systems, entanglement entropy \cite{refael,laflorencie2005,hoyos2007,devakul} and other entanglement measures \cite{fagotti,ruggiero,shapourian,rzj} have also been thoroughly studied in the ground state of random spin chains. In random antiferromagnetic spin chains, the strong-disorder renormalization group (SDRG) method is a powerful and insightful method to calculate low-energy properties \cite{mdh,fisherxx,im}. According to the iterative scheme of the SDRG procedure, which is approximative but becomes asymptotically exact at large scales, spins connected by strong couplings pair up to singlets, producing the so called random singlet state at zero temperature. This state, which is an approximation of the true (much more complicated) ground state, is a product state of independent spin singlets, and, due to its simplicity, it is analytically tractable. Applying this method in critical antiferromagnetic XXZ chains having a random singlet ground state, the average number of singlet bonds forming over a given point was calculated by Refael and Moore in the scaling limit as a function of the renormalization energy scale \cite{refael}. Using this, the average entanglement entropy of a block of $\ell$ contiguous spins was obtained to increase in leading order logarithmically as 
\be 
S_{\ell}=\frac{c_{\rm eff}}{3}\ln\ell + {\rm const},
\label{Sl}
\ee 
with the so-called effective central charge $c_{\rm eff}=\ln 2$. The validity of this approach was confirmed by a numerically exact diagonalization of the XX chain, which can be mapped to free fermions \cite{laflorencie2005}.
The average entanglement entropy in the random-singlet approximation is closely related to the distribution of singlet bond lengths $p_s(l)\simeq a_2l^{-2}$, the prefactor $a_2$ (times $\ln 2$) appearing in front of the logarithm in Eq. (\ref{Sl}) \cite{aperiodic,hoyos2007}.    

Besides the leading $l$-dependence, however, one might also be interested in the next-to-leading term in the finite-size scaling of the mean entanglement entropy. Information on the form of the correction term can be useful, in general, for numerical analyses in models in which the leading dependence is logarithmic but the effective central charge is a priori unknown. Such systems are, among others, the fermionic or spin chains with long-range interactions \cite{mohdeb}, aperiodic or quasiperiodic modulations \cite{aperiodic,rzj}, defects \cite{eisler,extended}, or junctions \cite{junction}.
In order to estimate the effective central charge in such cases, an extrapolation of the data obtained in finite systems must be performed, which necessitates an assumption about the form of corrections. 
To this end, we address the question of what forms the subleading term in the scaling of entanglement entropy as well as the singlet-length distribution have in the well-studied cases of random antiferromagnetic spin chains possessing a random singlet ground state. We mention that this question in homogeneous spin chains has been thoroughly studied \cite{igloi_lin,nienhuis,essler}.    
The SDRG calculations providing the leading logarithmic dependence on the block size rely on fixed-point solutions in the scaling limit in which the terms carrying information on the initial distribution of randomness are omitted. To obtain the corrections, however, this approach is insufficient. Nevertheless, it can be achieved, at least within the SDRG approach, by considering a special solution of power-law form of the SDRG flow equation which is valid from the beginning and changes through a single parameter along the SDRG trajectory \cite{igloi2002,hoyos2007}. Besides analytical and numerical SDRG calculations, we shall also present how the singlet lengths defined within the SDRG approach can be interpreted and numerically calculated in the XX chain by exact diagonalization of the equivalent free-fermion model. 

We emphasize that, although the main focus of this work is on the entanglement entropy in the ground state of antiferromagnetic spin chains, the results obtained here are more generic in several respects. As it will be discussed in detail in section \ref{sec:discussion}, they have implications also on correlation functions, are expected to be valid also for excited states, and for other models like the random transverse-field Ising chain, as well.   

The paper is organized as follows. The model and its SDRG scheme is described in section \ref{sec:model}. The special case of the XX chain is analyzed by the SDRG method in section \ref{sec:sdrg}. Here, earlier results necessary for our calculations are recapitulated, then an analytical calculation of the leading correction to the singlet-length distribution is presented, and from this, the correction to the entanglement entropy is inferred. In section \ref{sec:numerical}, numerical results obtained by the SDRG method and by exact diagonalization for the XX chain are presented. Results are discussed in section \ref{sec:discussion}. 

%%%%%%%%%%%%%%%%%%%%%%%%%%%%%%%%%%%%%%%%%%%%%%%%%%%%%%%%%%%%%%%%%%%%%%%%%%%%%%
%%%%%%%%%%%%%%%%%%%%%%%%%%%%%%%%%%%%%%%%%%%%%%%%%%%%%%%%%%%%%%%%%%%%%%%%%%%%%%
\section{The model}
\label{sec:model}

In this work, we consider random, antiferromagnetic spin-$\frac{1}{2}$ XXZ chains having the Hamiltonian:
\be 
H=\sum_nJ_n(S_n^xS_{n+1}^x+S_n^yS_{n+1}^y+\Delta S_n^zS_{n+1}^z),
\ee
where $S_n^{x,y,z}$ are spin operators at site $n$, the $J_n$ are positive, independent, identically distributed, random couplings, and the anisotropy parameter $\Delta$ is restricted to $0\le\Delta\le 1$. 
This model is known to have a random singlet ground state, and the SDRG scheme is summarized as follows \cite{mdh,fisherxx}. 
The pair of spins connected by the largest coupling $\Omega=\max_n\{J_n\}$ is picked, projected to a singlet state $\frac{1}{\sqrt{2}}(|\uparrow\downarrow\rangle-|\downarrow\uparrow\rangle)$ and removed from the chain, while an effective bond between spins neighboring to the decimated pair is created. 
Labeling the four spins involved in this procedure by $1$,$2$,$3$, and $4$, the coupling and anisotropy parameter of the newly formed bond (between spins $1$ and $4$) are calculated by second-order perturbation theory to be: 
\be
\tilde J=\frac{J_{1}J_{3}}{(1+\Delta_2)\Omega}, \qquad 
\tilde \Delta=\frac{1+\Delta_2}{2}\Delta_{1}\Delta_{3}. 
\ee
The length of the new bond will simply be
\be
\tilde l=l_1+l_2+l_3.
\ee
We assume that the bonds are initially of unit length, $l_n=1$, leading to that the renormalized bond lengths are all odd integers. During the SDRG procedure, the couplings become correlated with the bond lengths but the parameters of different bonds remain uncorrelated.

\section{SDRG analysis of the XX chain}
\label{sec:sdrg}

\subsection{Preliminaries}

First, we will consider the special case of the XX model, $\Delta=0$, for which analytical results of the SDRG approach exist.
The following master equation can be formulated for the joint probability distribution $P_{\Omega}(J,l)$ of couplings and bond lengths at energy scale $\Omega$: 
\beqn 
\frac{\partial P_{\Omega}(J,l)}{\partial\Omega}=-\int dJ_1dJ_3\sum_{l_1,l_2,l_3}P_{\Omega}(J_1,l_1)P_{\Omega}(\Omega,l_2)P_{\Omega}(J_3,l_3)\times \nonumber \\
\times\delta_{l,l_1+l_2+l_3}\delta(J-\frac{J_1J_3}{\Omega}). \nonumber
\\
\label{master}
\eeqn   
Using the generating function $\hat P_{\Omega}(J,\lambda)=\sum_le^{-\lambda l}P_{\Omega}(J,l)$, which was formally written as a Laplace transform with respect to a continuous variable $l$ in Refs. \cite{igloi2002,hoyos2007}, the master equation  assumes a simpler form:
\beqn 
\frac{\partial\hat P_{\Omega}(J,\lambda)}{\partial\Omega}=-\hat P_{\Omega}(\Omega,\lambda)\times \nonumber \\
\times\int dJ_1dJ_3\hat P_{\Omega}(J_1,\lambda)\hat P_{\Omega}(J_3,\lambda)\delta(J-\frac{J_1J_3}{\Omega}).
\eeqn    
We consider the special solution of this equation found in the closely related SDRG scheme of the transverse-field Ising chain \cite{igloi2002} and adapted to the XX chain as \cite{hoyos2007}: 
\be 
\hat P_{\Omega}(J,\lambda)=\frac{\alpha_{\Omega}(\lambda)}{\Omega}\left(\frac{\Omega}{J}\right)^{1-\beta_{\Omega}(\lambda)},
\label{special}
\ee
where 
\be 
\alpha_{\Omega}(\lambda)=
\frac{\sqrt{\theta_0^2-c^2}}{\cosh(c\Gamma)+\frac{\theta_0}{c}\sinh(c\Gamma)}.
\label{alpha}
\ee
Here, $\Gamma$ is a logarithmic energy scale, $\Gamma=\ln(\Omega_0/\Omega)$, where $\Omega_0$ is the initial value of $\Omega$; the function $c=c(\lambda)$ is independent of $\Omega$ and has the form
\be 
c(\lambda)=\theta_0\sqrt{1-e^{-2\lambda}},
\label{clambda}
\ee
whereas the function $\beta_{\Omega}(\lambda)$, the concrete form of which is not needed in the sequel, is given by $c^2=\beta^2_{\Omega}-\alpha_{\Omega}^2$. 
The special solution in Eq. (\ref{special}) corresponds to a power-law marginal distribution of couplings
\be
\hat P_{\Omega}(J,\lambda=0)=P_{\Omega}(J)=\frac{\theta(\Omega)}{\Omega}\left(\frac{\Omega}{J}\right)^{1-\theta(\Omega)},
\ee
with $\theta(\Omega)=(\theta_0^{-1}+\Gamma)^{-1}$, $\theta_0$ denoting the exponent of the initial distribution of couplings at $\Gamma=0$.  
The special solution tends to the fixed-point solution found by Fisher in the limit $\Omega\to 0$, which is attractive for any sufficiently regular initial distribution of couplings \cite{fisherxx}. 
Along this special trajectory, the fraction of active (not yet decimated) spins varies with the energy scale as 
\be 
n_{\Gamma}=\frac{1}{(1+\theta_0\Gamma)^2}.
\label{ngamma}
\ee 

\subsection{Distribution of singlet lengths}

First, we consider the distribution $p_s(l)$ of singlet bond lengths. This can be calculated by collecting the newly formed singlet lengths along the trajectory all the way to the fixed point at $\Omega=0$ as \cite{hoyos2007}: 
\be 
p_s(l)=2\int_0^{\Omega_0}n_{\Gamma}P_{\Omega}(\Omega,l)d\Omega.
\label{ps_int}
\ee
In principle, the distribution of singlet lengths could be obtained by reconstructing $P_{\Omega}(\Omega,l)$ from the generating function $\hat P_{\Omega}(\Omega,\lambda)=\alpha_{\Omega}(\lambda)/\Omega$, and performing the integration in Eq. (\ref{ps_int}). 
Essentially this way was followed in Ref. \cite{hoyos2007}, by taking the scaling limit $\lambda\to 0$, $\Gamma\to\infty$ so that $\lambda^{1/2}\Gamma\sim c\Gamma=O(1)$, which amounts to neglecting the term $\cosh(c\Gamma)$ in the denominator in Eq. (\ref{alpha}). Inverse Laplace transforming $\alpha_{\Omega}(\lambda)$ in this scaling limit leads ultimately to $p_s(l)=\frac{2}{3}l^{-2}[1+O(1/\sqrt{l})]$. The leading term obtained in this way agrees with the one that can be inferred from the leading block-size dependence of the mean entanglement entropy in Eq. (\ref{Sl}) calculated in Ref. \cite{refael}. But, the usage of this scaling limit is not expected to correctly account for the subleading terms which are affected by the full RG trajectory, not only by the close vicinity of the fixed point. We can see that the denominator in Eq. (\ref{alpha}) in the limit $c\to 0$ ($\lambda\to 0$) tends to $1+\theta_0\Gamma$, but, in the scaling limit, only $\theta_0\Gamma$ remains.  
Our numerical results obtained by the SDRG method (see later) indicate, however, that the correction is weaker, being in the order of $1/l$, so the correction $O(1/\sqrt{l})$ must be an artifact of the simplification in the scaling limit.  To improve this way of calculation so to obtain the correct subleading term seems to be hard, therefore we followed a different route.

For the generating function $\hat p_s(\lambda)=\sum_le^{-\lambda l}p_s(l)$, we obtain from Eqs. (\ref{ps_int}) and (\ref{special}) the form: 
\be
\hat p_s(\lambda)=2\int_0^{\infty}n_{\Gamma}\alpha_{\Omega}(\lambda)d\Gamma,
\label{pgen}
\ee
which is written as an integral over $\Gamma$. 
Guided by the numerical SDRG results, we assume the following expansion of the singlet-length distribution for $l\gg 1$: 
\be 
p^0_s(l)\simeq\sum_{n=2}^{\infty}\frac{a_n}{l^n}
\label{pseries}
\ee
for odd $l$, and $p^0_s(l)=0$ for even $l$, where $a_2,a_3,\dots$ are constants and, from previous results, we know only $a_2=2/3$.  
The generating function of such a fat-tailed distribution like $p^0_s(l)$ is nonanalytic at $\lambda=0$, containing a contribution proportional to $\ln\lambda$. To see this, we calculate the derivative of the generating function of the first term of the series, $\frac{d}{d\lambda}\sum_le^{-\lambda l}\frac{a_2}{l^2}=-\sum_le^{-\lambda l}\frac{a_2}{l}$, where the summation goes over odd integers only. This series sums up to $\frac{a_2}{2}\ln\frac{1-e^{-\lambda}}{1+e^{-\lambda}}=\frac{a_2}{2}\ln\frac{\lambda}{2}+O(\lambda^2)$. 
Similarly, we can see that the $n-1$st derivative of the generating function of the 
 $n$th term of $p^0_s(l)$ gives $(-1)^n\frac{a_n}{2}\ln\frac{\lambda}{2}+O(\lambda^2)$. 
Then, by integration, we obtain that the generating function 
$\hat p^0_s(\lambda)=\sum_le^{-\lambda l}p^0_s(l)$ must contain a singular part which is proportional to $\ln\lambda$ for $\lambda\to 0$:
\be 
[\hat p^0_s(\lambda)]_{\rm singular}=\sum_{n=2}^{\infty}(-1)^n\frac{a_n}{2}\frac{\lambda^{n-1}}{(n-1)!}\ln\lambda.
\label{singular}
\ee
Now, we try to identify such singular terms in the generating function represented as an integral in Eq. (\ref{pgen}). First, let us inspect the integrand in Eq. (\ref{pgen}). The factor $n_{\Gamma}$ is proportional to $\Gamma^{-2}$ for large $\Gamma$, while the second factor is $\alpha_{\Omega}(\lambda)\sim ce^{-c\Gamma}$ for $\Gamma\gg c^{-1}$.
Since we are looking for a contribution of the integral proportional to $\ln c$ in the limit $c\to 0$, we can cut off the integration at $\Gamma=c^{-1}$ rather than performing it to infinity; the dropped part of the integral is $O(c^2)$ and non-singular, therefore uninteresting from our point of view. 
Denoting the denominator of $\alpha_{\Omega}(\lambda)$ in Eq. (\ref{alpha}) by 
$R(c_0,\Gamma_0)\equiv\cosh(c_0\Gamma_0)+c_0^{-1}\sinh(c_0\Gamma_0)$, where $c_0\equiv c/\theta_0$ and $\Gamma_0\equiv \theta_0\Gamma$ we can recast $\alpha_{\Omega}(\lambda)$ as 
\be
\alpha_{\Omega}(\lambda)=
\frac{\theta_0\sqrt{1-c_0^2}}{1+\Gamma_0}\frac{1}{1+ [(1+\Gamma_0)^{-1}R-1]}.
\label{alpha2}
\ee      
Then we apply the series expansion of $\cosh(x)$ and $\sinh(x)$ functions around $x=0$ to obtain $T\equiv (1+\Gamma_0)^{-1}R-1=\frac{1}{1+\Gamma_0}\sum_{n=2}^{\infty}\frac{b_n(c_0\Gamma_0)^n}{n!}$, where $b_n$ is $1$ for $n$ even and $c_0^{-1}$ for $n$ odd. Since $0<T<1$ in the domain $0\le\Gamma_0\le c_0^{-1}$ for any $c_0>0$, we can expand the second factor in Eq. (\ref{alpha2}) in a geometric series 
\be
\frac{1}{1+T}=\sum_{n=0}^{\infty}({-1})^n(1+\Gamma_0)^{-n}\sum_{k_1,k_2,\dots,k_n=2}^{\infty}\frac{c_0^{s_n-o_n}\Gamma_0^{s_n}}{k_1!k_2!\dots k_n!},
\ee  
where $s_n=\sum_{i=1}^nk_i$ and $o_n$ is the number odd indices $k_i$. 
Using this expansion and Eq. (\ref{ngamma}), the integrand can be written as a series with both negative and positive powers of $\Gamma_0+1$. Clearly, integrating from $0$ to $1/c_0$ a contribution proportional to $\ln c_0$ will arise from the terms proportional to $(1+\Gamma_0)^{-1}$. Collecting these terms, we obtain ultimately for the singular part: 
\beqn
[\hat p_s(\lambda)]_{\rm singular}=-2\sqrt{1-c_0^2}\times \nonumber \\ 
\times\sum_{n=1}^{\infty}({-1})^n\sum_{k_1,k_2,\dots,k_n=2}^{\infty}(-1)^{s_n}\binom{s_n}{n+2}\frac{c_0^{s_n-o_n}}{k_1!k_2!\dots k_n!}\ln c_0. \nonumber \\
\label{sing_result}
\eeqn   
We can see that the expression in front of $\ln c_0$ contains only even powers of $c_0$, and using that $c_0^2=1-e^{-2\lambda}$ it is therefore analytic in $\lambda$, in accordance with the assumption in Eq. (\ref{pseries}) leading to Eq. (\ref{singular}).
Evaluating Eq. (\ref{sing_result}) for the first few orders of $c_0$, we obtain:
\be
[\hat p_s(\lambda)]_{\rm singular}=\left[\frac{1}{3}c_0^2+O(c_0^8)\right]\ln c_0.
\label{pc0}
\ee
Substituting $c_0^2=-\sum_{n=1}^{\infty}\frac{(-2\lambda)^n}{n!}$ [obtained from Eq. (\ref{clambda})] into Eq. (\ref{pc0}) and comparing it with Eq. (\ref{singular}), we obtain for the first three expansion coefficients: 
\be 
a_2=\frac{2}{3}, \quad a_3=\frac{4}{3}, \quad a_4=\frac{8}{3}.
\label{coeffs}
\ee  
The coefficient $a_2=2/3$ of the leading term agrees with previous results of Refs. \cite{refael,hoyos2007}.

\subsection{Entanglement entropy}

Next, we consider the scaling of the entanglement entropy within the SDRG approach. First, we count the singlets crossing a wall which divides the chain into two semi-infinite parts. The mean number of such singlets created down to energy scale $\Omega$ is denoted by $N(\Omega)$. This quantity was calculated in two different ways in the scaling limit in Refs. \cite{refael,hoyos2007}. From $N(\Omega)$, the leading block-size dependence of the mean entanglement entropy can be inferred by using the relationship between the mean bond length $l=1/n_{\Gamma}$ and $\Gamma$ given in Eq. (\ref{ngamma}).   
Here, we use the exact starting point for the calculation of $N(\Omega)$ formulated in Ref. \cite{hoyos2007} in terms of the distribution $P_{\Omega}(\Omega,l)$ of singlet lengths created at scale $\Omega$: 
\be 
N(\Omega)=\int_{\Omega}^{\Omega_0}d\Omega n_{\Gamma}\sum_lP_{\Omega}(\Omega,l)l.
\label{N}
\ee
In Ref. \cite{hoyos2007}, this formula was evaluated using the explicit form of the distribution $P_{\Omega}(\Omega,l)$ obtained in the scaling limit. This yields 
$N(\Omega)=\frac{1}{3}[\ln(1+\Gamma_0)-1+(1+\Gamma_0)^{-1}]$, and through the substitution $\ell=1/n_{\Gamma}=(1+\Gamma_0)^2$ leads to the form in Eq. (\ref{Sl}) with an additional $O(\ell^{-1/2})$ correction term \cite{hoyos2007}. 
Now, we evaluate Eq. (\ref{N}) exactly along the special trajectory, using that the expected value of singlet lengths at scale $\Omega$ can directly be calculated from the generating function through:
\beqn 
\sum_lP_{\Omega}(\Omega,l)l&=&-\frac{d\hat P_{\Omega}(\Omega,\lambda)}{d\lambda}|_{\lambda=0}= 
-\frac{1}{\Omega}\frac{d\alpha_{\Omega}(\lambda)}{d\lambda}|_{\lambda=0}= \nonumber \\
&=&-\frac{2}{\Omega}\frac{d\alpha_{\Omega}}{dc_0^2}|_{c_0^2=0}.
\eeqn
Then, from Eq. (\ref{alpha}) we obtain 
\be 
\sum_lP_{\Omega}(\Omega,l)l=\frac{\theta_0}{\Omega}\frac{2}{3}\left[\frac{1}{2}(1+\Gamma_0)+\frac{1}{(1+\Gamma_0)^2}\right].
\ee
Substituting this expression into Eq. (\ref{N}) and evaluating the integral we find that
\be
N(\Omega)=\frac{1}{3}\ln(1+\Gamma_0) + \frac{2}{9}\left[1-\frac{1}{(1+\Gamma_0)^{3}}\right].
\ee
Thus, the subleading term decreases as $\sim \Gamma^{-3}$ for large $\Gamma$, and the substitution $\ell=1/n_{\Gamma}=(1+\Gamma_0)^2$ then results in a correction of $O(\ell^{-3/2})$. Nevertheless this transformation of $\Gamma$-dependence to $\ell$-dependence, which certainly provides the correct leading-order behavior, has to be taken cautiously if one is interested in the subleading term, as we can see later. 

The previous approach counts the number of crossing singlets during the SDRG procedure as a function of the renormalization scale $\Gamma$. Alternatively, one can write the mean entanglement entropy $S_{\ell}$ of a block of $\ell$ contiguous sites (embedded in an infinite chain) in a straightforward way in terms of the distribution of singlet lengths $p_s(l)$ in the random singlet phase as 
\be 
\frac{S_{\ell}}{\ln 2}=\sum_{l<\ell}p_s(l)l + \ell\sum_{l\ge \ell}p_s(l).
\label{Ssum}
\ee
Here, the expression on the r.h.s. is the mean number of singlets with exactly one constituent spin within the block, and we have used that each such singlet gives an independent contribution of $\ln 2$ to the entanglement entropy.  

Let us consider the first sum in Eq. (\ref{Ssum}). The leading term of the distribution $\frac{a_2}{l^2}$ provides the leading logarithmic divergence of the entanglement entropy with an $O(\ell^{-1})$ correction: 
$\frac{a_2}{2}\ln \ell+\frac{a_2}{2}(\gamma+\ln 2)+O(\ell^{-1})$, where $\gamma$ denotes the Euler-Mascheroni constant. The subleading term of $p_s(l)$ brings another correction term of the same order: $a_3\frac{\pi^2}{8}-\frac{a_3}{2}\frac{1}{\ell}+O(\ell^{-2})$. 
The second sum in Eq. (\ref{Ssum}) gives a contribution $\frac{a_2}{2}+O(\ell^{-1})$, where the correction is composed of both the contributions of the first and second term of the expansion of $p_s(l)$.
We thus conclude that the subleading term in the size-dependence of the entanglement entropy is $O(\ell^{-1})$: 
\be 
\frac{S_{\ell}}{\ln 2}=\frac{1}{3}\ln \ell + {\rm const} + O(\ell^{-1}),
\label{Scorr}
\ee
where the correction is determined by the first two terms of the expansion of singlet-length distribution given in Eq. (\ref{pseries}). 
According to the analytical results obtained for the coefficients in Eq. (\ref{coeffs}), the coefficient of the $O(\ell^{-1})$ correction term is independent of the parameter $\theta_0$ of the special solution. 

\section{Numerical results}
\label{sec:numerical}

\subsection{Distribution of singlet lengths}

\begin{figure}
\begin{center}
\includegraphics[width=80mm, angle=0]{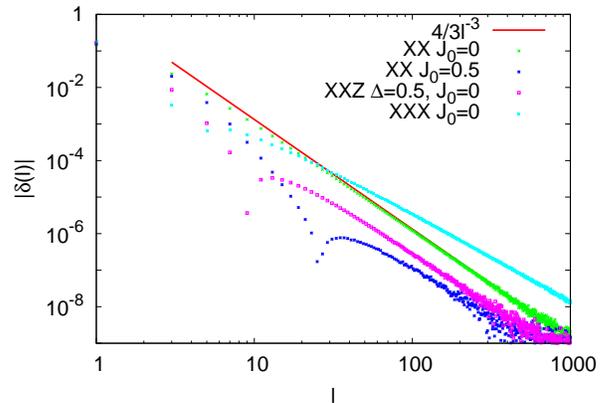}
%ccorr2.ps
\end{center} 
\caption{
\label{fig_rg_corr}
The magnitude of the correction $\delta(l)=p_s(l)-\frac{2}{3}l^{-2}$ of the singlet-length distribution obtained by the numerical SDRG method for different variants of the random XXZ chain. The solid line is the analytically obtained prediction for the subleading term valid for the case $J_0=0$, $\Delta=0$. 
}
\end{figure}

\begin{figure}
\begin{center}
\includegraphics[width=80mm, angle=0]{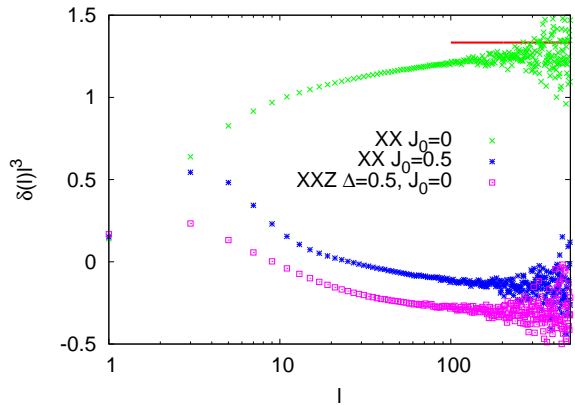}
%ccorr2+.ps
\end{center} 
\caption{
\label{fig_rg_coeff}
The scaled correction of the singlet-length distribution, $\delta(l)l^3$, which is expected to tend to the prefactor $a_3$ of the subleading term. The horizontal line indicates the analytically obtained limiting value $4/3$ valid for the XX model with $J_0=0$. 
}
\end{figure}

\begin{figure}
\begin{center}
\includegraphics[width=80mm, angle=0]{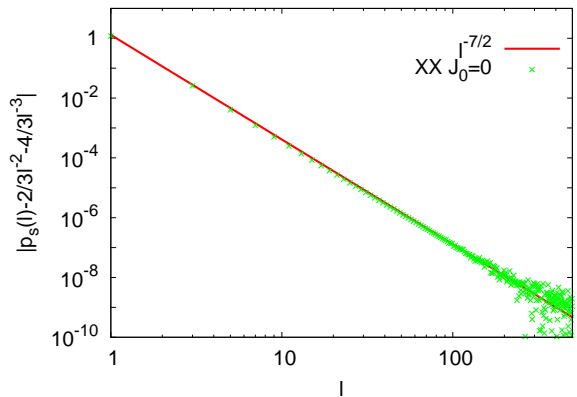}
%ccorr4.ps
\end{center} 
\caption{
\label{fig_corr_half}
The correction of the singlet-length distribution beyond the subleading term for the  XX model with $J_0=0$. The data are obtained by the numerical SDRG method. The solid line corresponds to an algebraic decrease proportional to $l^{-7/2}$.
}
\end{figure}

In order to check the results obtained in the previous section, we calculated the distribution of singlet lengths by numerically implementing the SDRG method. We considered uniform distributions of couplings with the support $[J_0,J_0+1]$. For the XX model ($\Delta=0$), $J_0=0$ corresponds to the initial point of the special trajectory with $\theta_0=1$, whereas for $J_0>0$, it does not belong to the initial domain of the special solution. We considered periodic chains with $L=10^5$ sites, and defined the distance between two sites $i$ and $j$ as $l=\min\{|i-j|,L-|i-j|\}$. The distribution $p_s(l)$ was determined from data obtained in $10^7$ independent random samples. 
We calculated the correction to the leading term of the distribution 
\be 
\delta(l)=p_s(l)-\frac{2}{3}l^{-2}
\ee 
and plotted it for different values of $J_0$ and $\Delta$ in Figs. \ref{fig_rg_corr} and \ref{fig_rg_coeff}. 
%%%%%%%%%%%%%%%%%%%%%%%%%%%%%%%%%%%%%%%%%%%%%%%%%%%%%%%%%%%%%%%%%%%%%%%%%%%%%%%
%%%%%%%%%%%%%%%%%%%%%%%%%%%%%%%%%%%%%%%%%%%%%%%%%%%%%%%%%%%%%%%%%%%%%%%%%%%%%
%%%%%%%%%%%%%%%%%%%%%%%%%%%%%%%%%%%%%%%%%%%%%%%%%%%%%%%%%%%%%%%%%%%%%%%%%%%%%%%
%%%%%%%%%%%%%%%%%%%%%%%%%%%%%%%%%%%%%%%%%%%%%%%%%%%%%%%%%%%%%%%%%%%%%%%%%%%%%
For the special solution of the XX chain ($J_0=0$, $\Delta=0$), the correction is $\delta(l)=\frac{4}{3}l^{-3}+O(l^{-4})$ according to the analytical results. As can be seen in Fig. \ref{fig_rg_coeff}, the form of the subleading term $\frac{4}{3}l^{-3}$ is consistent with the numerical data, $\delta(l)l^3$ tending indeed toward $4/3$. Nevertheless the approach is much slower than that obtained from the analytical calculations $\delta(l)l^3=\frac{4}{3}+O(l^{-1})$, and even the sign of the correction of $\delta(l)l^3$ is opposite (negative). This shows that a correction of the form other than that assumed in Eq. (\ref{pseries}) must also be present, which mask the term $O(l^{-4})$. In Fig. \ref{fig_corr_half}, we find that the correction next to the subleading term is proportional to $l^{-7/2}$.  
%%%%%%%%%%%%%%%%%%%%%%%%%%%%%%%%%%%%%%%%%%%%%%%%%%%%%%%%%%%%%%%%%%%%%%%%%%%%%%%
%%%%%%%%%%%%%%%%%%%%%%%%%%%%%%%%%%%%%%%%%%%%%%%%%%%%%%%%%%%%%%%%%%%%%%%%%%%%%

For the distribution with $J_0=0.5$ in the XX case and for the XXZ model with $\Delta=0.5$ and $J_0=0$, the numerical results still show a subleading term of $O(l^{-3})$ but the coefficient of this term differs from $4/3$ and is non-universal. This is expected to be the case generally for $\Delta<1$, where the renormalization flow is attracted by the XX fixed point \cite{fisherxx}.   

\begin{figure}
\begin{center}
\includegraphics[width=80mm, angle=0]{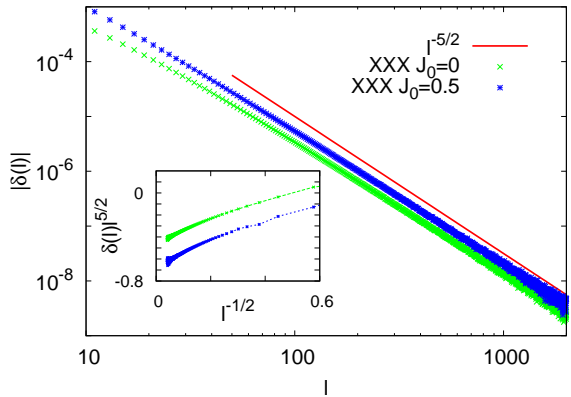}
%ccorrXXX.ps
\end{center} 
\caption{
\label{fig_rg_XXX} Correction of the singlet-length distribution in random XXX chains, obtained by the numerical SDRG method. The inset shows the rescaled correction $\delta(l)l^{5/2}$ plotted against $l^{-1/2}$. 
}
\end{figure}

For the isotropic case $\Delta=1$, for which the flow tends to the Heisenberg fixed point \cite{fisherxx}, the numerical results show a different form of the subleading term. As can be seen in Fig. \ref{fig_rg_XXX}, the corrections are stronger than those found at the XX fixed point, being roughly proportional to $l^{-5/2}$. Plotting the rescaled correction $\delta(l)l^{5/2}$ against $l^{-1/2}$ (see the inset of the figure), the asymptotically linear behavior indicates a further correction term of order $l^{-3}$.
%%%%%%%%%%%%%%%%%%%%%%%%%%%%%%%%%%%%%%%%%%%%%%%%%%%%%%%%%%%%%%%%%%%%%%%%%%%%%%%
%%%%%%%%%%%%%%%%%%%%%%%%%%%%%%%%%%%%%%%%%%%%%%%%%%%%%%%%%%%%%%%%%%%%%%%%%%%%%

\begin{figure}
\begin{center}
\includegraphics[width=80mm, angle=0]{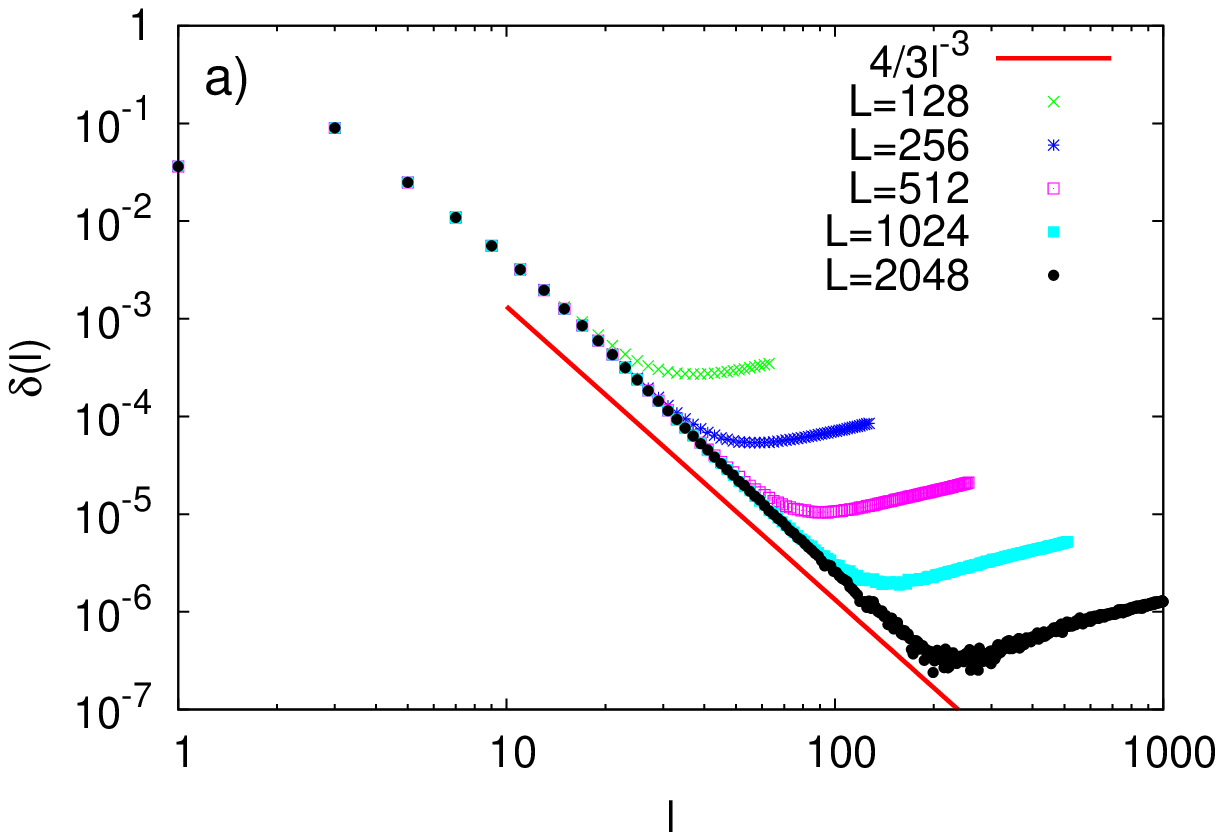}
\includegraphics[width=80mm, angle=0]{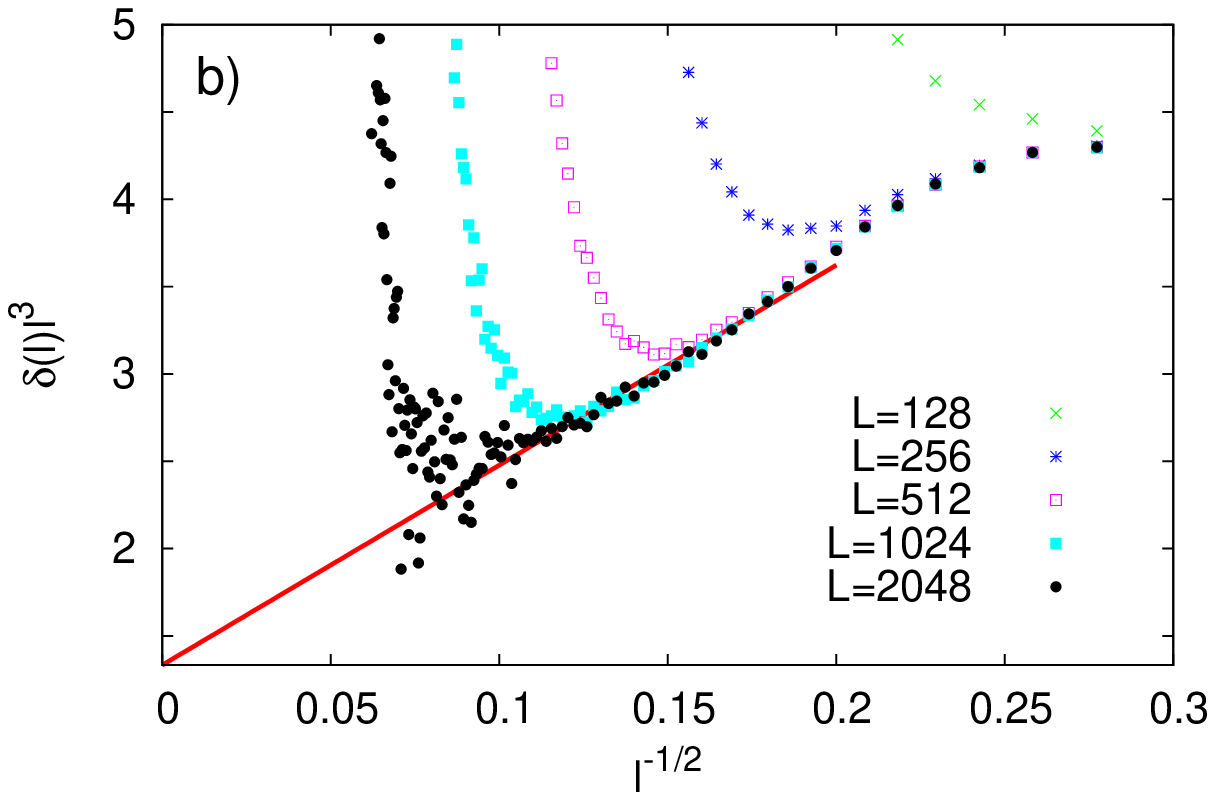}
%ldsr3.ps
%ldsr3++.ps
\end{center} 
\caption{
a) Correction to the distribution of distances between odd and even sublattice maxima of the one-particle states in the free-fermion model equivalent to the XX chain. The data were obtained by an exact diagonalization of the free-fermion Hamiltonian for different system sizes $L$. The straight line represents the subleading term obtained by the SDRG approach.   
b) The scaled correction, $\delta(l)l^3$, plotted against $O(l^{-1/2})$.  The straight line having an intercept $4/3$ is a guide to the eye.
\label{fig_ED_pl}
}
\end{figure}

For the XX model, which can be mapped to free fermions by the Jordan-Wigner transformation, we can confront the SDRG results with those obtained by exact diagonalization. To do so, the singlet lengths, which are natural in the SDRG approach, have to be interpreted in the original model. 
For this, we use the free-fermion representation of the XX chain 
\be 
H_f=\frac{1}{2}\sum_nJ_n(c_n^{\dagger}c_{n+1}+c_{n+1}^{\dagger}c_{n}),
\ee
where $c_n$ and $c_n^{\dagger}$ are fermion annihilating and creating operators, respectively, at site $n$. We assume that the number of sites is even, so that the lattice is bipartite, and, due to this, the fermion system has a sublattice symmetry. As a consequence, the one-particle energies appear in pairs $\pm\epsilon_k$, and the ground state is at half-filling. 
In this fermion system, the SDRG approximation means that the one-particle states are perfectly localized on two sites, $i$ and $j$, which belong to different sublattices. For negative energies, the $k$th eigenstate is thus of the form
\be 
\psi^{SDRG}_k(n)=\frac{1}{\sqrt{2}}(\delta_{in}-\delta_{jn}).
\ee
Clearly, the spin singlets of the XX chain correspond to the one-particle states of the fermion system. 
The true eigenstates $\psi_k(n)$ are, although not perfectly, but still localized around the two sites picked by the SDRG. As a consequence of the bipartite structure, one can show at low energies $|\epsilon_k|\ll 1$ that, if the wave function is $O(1)$ on one of the sublattices in some region, then  $\psi_k(n)$ in the same region but on the other sublattice must be small. Thus the one-particle states are concentrated in two spatially separated regions hosted essentially on different sublattices around the two SDRG sites. 
As centers of these two regions for a given one-particle state $\psi_k$, we considered the sites $i_k$ and $j_k$ with the maximal magnitude of the wave function on the odd and the even sublattice, respectively:
\beqn 
|\psi_k(i_k)|&=&\max_n\{|\psi_k(2n+1)|\}  \nonumber \\
|\psi_k(j_k)|&=&\max_n\{|\psi_k(2n)|\}.         
\eeqn
By diagonalizing the Hamiltonian of the fermion model numerically up to a lattice size $L=2048$, we calculated the distance $l_k=\min\{|i_k-j_k|,L-|i_k-j_k|\}$ for each eigenstate, and repeating this calculation for $10^7$ random realizations of the disorder with $J_0=0$, we measured the distribution $p_f(l)$. 
As can be seen in Fig. \ref{fig_ED_pl}, the correction $\delta(l)=p_f(l)-\frac{2}{3}l^{-2}$ of the distribution tends to zero proportionally to $l^{-3}$. The scaled correction $\delta(l)l^3$ has, however, strong corrections at the sizes available numerically, and it is thus not possible to extract the coefficient $a_3$ reliably. Assuming a correction of order $l^{-7/2}$ next to the subleading term as was found by the SDRG method, the data allow for an asymptotic value of the coefficient $a_3=4/3$, see the extrapolation in Fig.  \ref{fig_ED_pl}b. 
Nevertheless, the sign of the correction term $O(l^{-7/2})$ is different from that found by the SDRG method, which indicates the presence of further corrections which are not captured by the SDRG approximation.   

%%%%%%%%%%%%%%%%%%%%%%%%%%%%%%%%%%%%%%%%%%%%%%%%%%%%%%%%%%%%%%%%%%%%%%%%%%%%%%%
%%%%%%%%%%%%%%%%%%%%%%%%%%%%%%%%%%%%%%%%%%%%%%%%%%%%%%%%%%%%%%%%%%%%%%%%%%%%%

\subsection{Entanglement entropy}

\begin{figure}
\begin{center}
\includegraphics[width=80mm, angle=0]{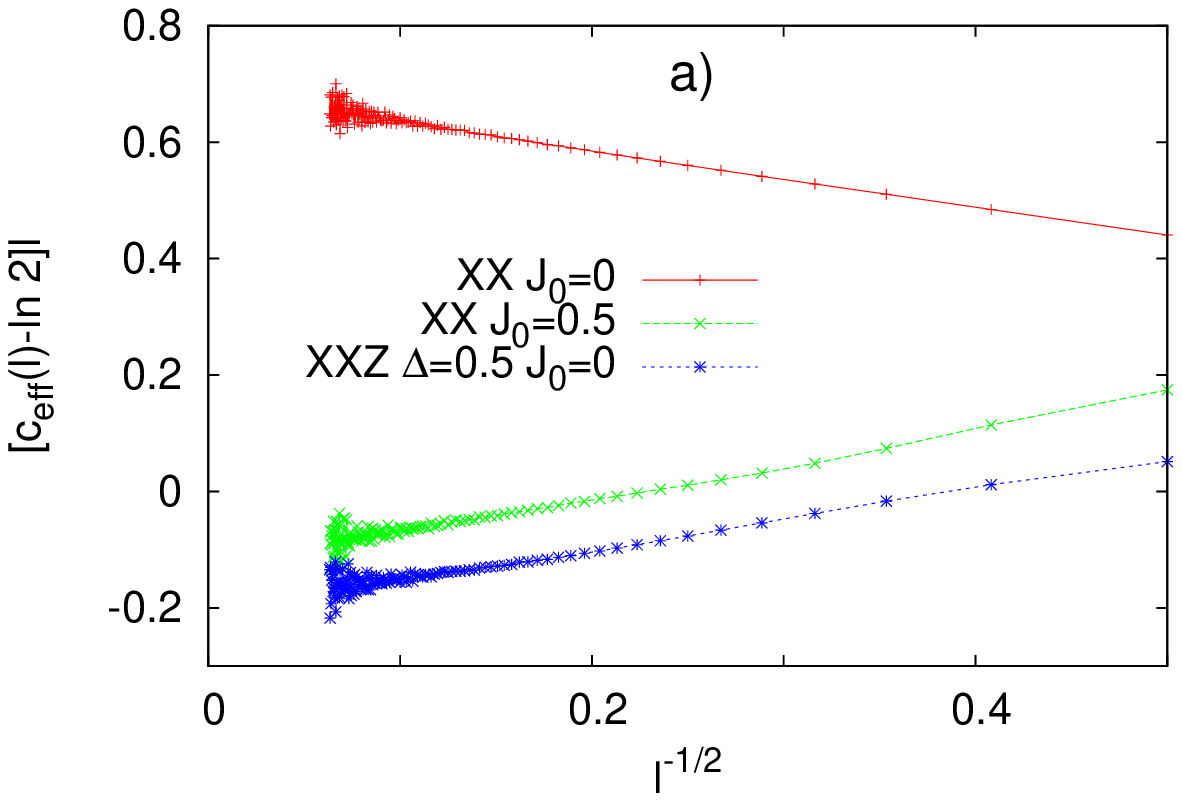}
\includegraphics[width=80mm, angle=0]{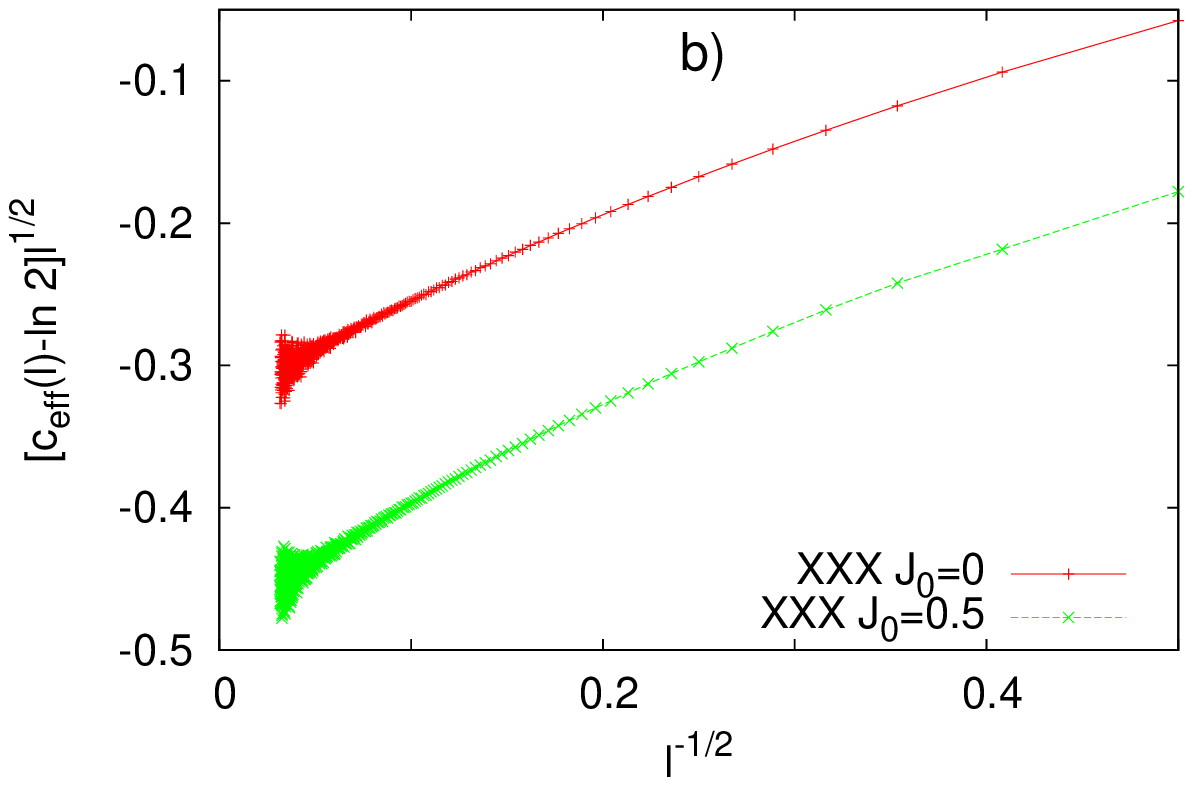}
%profsrpref+.ps
%profsrprefXXX+.ps
\end{center} 
\caption{
The scaled correction to the effective central charge, calculated numerically by the SDRG method for variants of random the XXZ chain (a) and the XXX chain (b).  
\label{fig_ceff}
}
\end{figure}

From the distribution of singlet lengths obtained numerically by the SDRG method we calculated the average entanglement entropy $S_{\ell}$ of subsystems of size $\ell$ through Eq. (\ref{Ssum}). In order to eliminate the constant term in the asymptotic dependence of $S_{\ell}$ on $\ell$, see Eq. (\ref{Scorr}), we considered the discrete derivative with respect to $\ln\ell$, a quantity which is practical for a numerical determination of $c_{\rm eff}$:
\be 
c_{\rm eff}(\ell)=3\frac{S_{\ell+s}-S_{\ell}}{\ln(\ell+s)-\ln\ell}.
\label{der}
\ee  
For $s=1$, this quantity shows an oscillatory behavior due to that the singlet lengths are odd, therefore we used $s=2$ in Eq. (\ref{der}).
From Eq. (\ref{Scorr}), we find then an $O(\ell^{-1})$ leading correction to the effective central charge:
\be 
c_{\rm eff}(\ell)=\ln 2 + O(\ell^{-1}) \qquad (\Delta<1)
\label{ceffperl}
\ee
which is expected to be valid for the XXZ chain with $\Delta<1$. 
Numerical results shown in Fig.~\ref{fig_ceff}a are compatible with this, and the coefficient of the correction term is found to be non-universal. Moreover, the linear asymptotic dependence of $[c_{\rm eff}(\ell)-\ln 2)]\ell$ on $\ell^{-1/2}$ indicates that the next correction term is $O(\ell^{-3/2})$.     
For the XXX chain ($\Delta=1$), the form of the correction is different from this. Using the numerically found subleading term of the distribution of singlet lengths, $\delta(\ell)\simeq b\ell^{-5/2}$, we obtain through Eq. (\ref{Ssum}), that the leading correction to the effective central charge is in the order of $\ell^{-1/2}$: 
\be 
c_{\rm eff}(\ell)\simeq\ln 2 + b\ln 2\ell^{-1/2} \qquad (\Delta=1).
\ee
The numerical results shown in Fig.~\ref{fig_ceff}b are in accordance with this form, although the convergence of $[c_{\rm eff}(\ell)-\ln 2]\ell^{1/2}$ to $b\ln2$ is slow due to the next correction term of order $\ell^{-1}$. 
%%%%%%%%%%%%%%%%%%%%%%%%%%%%%%%%%%%%%%%%%%%%%%%%%%%%%%%%%%%%%%%%%%%%%%%%%%%%%%%
%%%%%%%%%%%%%%%%%%%%%%%%%%%%%%%%%%%%%%%%%%%%%%%%%%%%%%%%%%%%%%%%%%%%%%%%%%%%%

\begin{figure}
\begin{center}
\includegraphics[width=8cm]{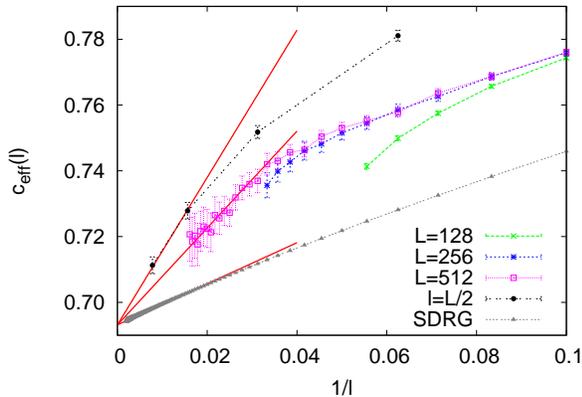}
%proftp2.ps
\end{center} 
\caption{
\label{fig_ED_ceff}
The effective central charge defined in Eq. (\ref{der}) and calculated by exact numerical diagonalization of the free-fermion model equivalent to the XX chain, as a function of the inverse subsystem size, for different system sizes $L$. The distribution of couplings was uniform with $J_0=0$. As a comparison, the data obtained by the SDRG method, as well as $c_{\rm eff}(L/2)$ calculated from the half-chain entropies [by setting $\ell=L/2$ and $s=L/2$ in Eq. (\ref{der})] are also shown. The straight lines have slopes $0.62$, $1.47$, and $2.24$.  
}
\end{figure}

We also calculated the entanglement entropy of the XX model by mapping it to free fermions and performing exact numerical diagonalization. The entanglement entropy can then be calculated from the eigenvalues of the correlation matrix $C_{ij}=\langle c_i^{\dagger}c_j\rangle$ restricted to the subsystem \cite{vidal,peschel}. We calculated the entanglement entropy of blocks with an even number of sites and averaged over $10^7$ random samples as well as for different positions of the block in the system. The size-dependent effective charges defined in Eq. (\ref{der}) are plotted in Fig. \ref{fig_ED_ceff}. 
The data are compatible with Eq. (\ref{ceffperl}), but the coefficient of the $O(\ell^{-1})$ correction term is found to be significantly greater than that calculated by the SDRG method. This shows again the presence of further corrections  related to the imperfect localization of states in the original model which the SDRG approximation can not account for. 

%%%%%%%%%%%%%%%%%%%%%%%%%%%%%%%%%%%%%%%%%%%%%%%%%%%%%%%%%%%%%%%%%%%%%%%%%%%%%%%
%%%%%%%%%%%%%%%%%%%%%%%%%%%%%%%%%%%%%%%%%%%%%%%%%%%%%%%%%%%%%%%%%%%%%%%%%%%%%

\section{Discussion}
\label{sec:discussion}

In this work, we studied the entanglement entropy of blocks of spins and the closely related distribution of singlet lengths in the random singlet phase of antiferromagnetic XXZ chains. The leading-order size-dependence of these quantities is well known by the strong-disorder renormalization group approximation. We addressed here a less studied question of how large the subleading terms are. 
Using a special trajectory of the SDRG flow valid for the XX chain, we found by an analytical calculation that the singlet-length distribution contains a series of integer powers of $1/l$, the leading term of which agrees with the well-known result $\frac{2}{3}l^{-2}$, see Refs. \cite{refael,hoyos2007}. Our numerical SDRG results confirm the form of the subleading term, $\frac{4}{3}l^{-3}$, and also reveal a term of order $l^{-7/2}$ which is not identified by our analytical method, and which masks the subsequent integer-power term. For general distributions of couplings not initiating the special trajectory and for the XXZ chain with an anisotropy parameter $0<\Delta<1$, we found by the numerical SDRG method that the subleading term is still proportional to $l^{-3}$ but the coefficient is non-universal.  
For the XX chain, we interpreted the concept of singlet lengths originating in the SDRG approach as the distance between maxima of the one-particle wave functions on the even and odd sublattices in the equivalent free-fermion model. 
We calculated the distribution of these distances by exact numerical diagonalization and found that the form of the subleading term is compatible with the SDRG prediction. Nevertheless, for a clearer conclusion on this issue, larger systems should be numerically investigated. 
For the isotropic XXX chain ($\Delta=1$), we found by the numerical SDRG method the correction to be stronger than for models attracted by the XX fixed point: the subleading term is proportional to $l^{-5/2}$ rather than to $l^{-3}$. 

Based on the numerical results, we conjecture that besides the analytically found integer powers of $1/l$ (starting with $l^{-2}$), the singlet-length distribution also comprises half-integer powers of $1/l$, starting with $l^{-7/2}$ for the XX fixed point and $l^{-5/2}$ for the XXX fixed point. Moreover, the coefficients apart from that of the leading term are non-universal.   

The subleading term of the singlet-length distribution appears also in the size-dependence of the average entanglement entropy and, for the XX fixed point, gives (together with other corrections of the same order) a subleading term of $O(\ell^{-1})$  with a non-universal coefficient next to the leading logarithmic divergence. This was confirmed by numerical SDRG results as well as by exact diagonalization for the XX chain. The true coefficient of the subleading term is found to be greater compared to that obtained by the SDRG approach. 
For the XXX chain, the subleading term of the singlet-length distribution which was found numerically to be of $O(\ell^{-5/2})$ dominates the other correction terms and leads to a subleading term of $O(\ell^{-1/2})$ next to the logarithmic divergence in the entanglement entropy. 

The leading term of the entanglement entropy scaling and the closely related singlet-length distribution are known to be universal and correctly captured by the SDRG approach. We have found in this work that, as opposed to this, the subleading term of these quantities is non-universal: although the exponent is universal for the XX fixed point, the coefficients depend on the form of coupling distribution and on the anisotropy parameter. Moreover, for the XXX fixed point even the exponent differs from that of the XX fixed point. 
We have also found that the SDRG approach, although properly gives the order of the subleading term of the entanglement entropy, fails to provide the correct coefficients. This failure must be related to the assumption on perfect localization of eigenstates which is at the core of the SDRG approach.       

Other quantities not studied in this work but which are closely related to the distribution of singlet lengths are the spin-spin correlation functions $C^{\alpha}(l)=\langle S_0^{\alpha}S_l^{\alpha}\rangle$ for $\alpha=x,y,z$. In the SDRG approach, their average value is directly proportional to the singlet-length distribution $p_s(l)$. In a recent exact diagonalization study of the correlation functions of the random XX chain, weak non-universal corrections have been seen in the transversal ($\alpha=z$) correlation function \cite{getelina}. It is an interesting question whether the corrections next to the leading term of the correlation function are compatible with the SDRG predictions of this work.
 
Besides the ground state, which we considered throughout this paper, the entanglement entropy of the random XX chain has also been studied in excited states sampled from a canonical ensemble in Ref. \cite{huang} by an extension of the SDRG method \cite{pekker}. Here, a leading logarithmic divergence of the entanglement entropy was found with a prefactor undetermined owing to degeneracies of the excited states \cite{huang}. Within the SDRG approach, the excited states have the same product state structure as the ground state has, with the difference that some of the spin pairs are in a triplet rather than a singlet state. Thus, the distribution of the length of spin pairs is the same, and, as a consequence, the $\ell$-dependent terms of the ground-state entanglement entropy, including the corrections found in this work, are modified in excited states by the same global (undetermined) factor.

It is worth mentioning that, as a consequence of the exact relationship between the entanglement entropy in the random XX chain and in the random transverse-field Ising chain \cite{ij}, the subleading term near the logarithmic one at the critical point of the latter model must be $O(\ell^{-1})$, as well. It would be desirable to give a quantitative description of the stronger corrections observed at the XXX fixed point within the SDRG approach and to confirm these results with numerical methods beyond the SDRG approach, which are left for future research.

\begin{acknowledgments}
The author thanks Ferenc Igl\'oi and Jos\'e A. Hoyos for useful discussions.
This work was supported by the National Research, Development and Innovation Office NKFIH under grant No. K128989. 
\end{acknowledgments}

\end{document}